# Knowledge Tectonics: A Geodynamic Inspired Framework for Modelling Epistemic Changes

Thomas Van Erven[1], Stelios Kontogiannis[2], Efstratios Kontopoulos[3,*], Konstantinos Avgerinakis[2], Andrea Scharnhorst[4] and Sándor Darányi[1]

[1] *Swedish School of Library and Information Science, University of Borås, Sweden*
[2] *Catalink Limited, Nicosia, Cyprus*
[3] *Choreograph (WPP Media), Copenhagen, Denmark*
[4] *DANS-KNAW, Royal Netherlands Academy of Arts and Sciences, The Hague, The Netherlands*

**Abstract**

Emerging knowledge can be envisioned as protruding magma. This paper describes how such a metaphoric view can be turned into a model to capture the "continental drift" of concepts in an epistemic lithosphere. We call this new approach *Knowledge Tectonics*. We detail conceptual, mathematical and engineering operations to create such a scalable framework which allows us to interpret and manage knowledge evolution within Semantic Web environments. We use the WikiArt Emotions dataset which contains information on 4,105 paintings spanning 600 years to construct a proof-of-concept interface which enables visual analytics of where artworks are situated in a specific landscape of features. We demonstrate how fused semantic and pragmatic metadata can be modelled as evolving pressure zones. This way we are making "forces" behind the evolution of artifacts of creativity visible. Core elements of our new workflow are gradient vector fields derived from Poisson potential surfaces applied onto dynamic knowledge graphs, eventually capturing stylistic and emotional shifts as directed intensity flows. By treating artefacts of creative processes as a dynamic manifold, we provide a novel methodology for quantifying the "drift" of human inquiry. We argue that this approach is applicable also to other areas of creative human actions, including scientific knowledge production.

**Keywords**

Knowledge Tectonics, Knowledge Graphs, Semantic Web, Vector-Augmented Knowledge Graphs, Evolving Epistemic Changes

## 1. Introduction

To understand the evolution of knowledge as visible in artefacts of human creativity (from objects of arts, of craftsmanship and engineering up to scientifically documented ideas) has been an intriguing topic for many disciplines. In the design of knowledge graphs how to best

* Corresponding author.
thomas.erven@gmail.com (T. van Erven); skontogiannis@catalink.eu (S. Kontogiannis); stratos.kontopoulos@choreograph.com (E. Kontopoulos); koafgeri@catalink.eu (K. Avgerinakis); andrea.scharnhorst@dans.knaw.nl (A. Scharnhorst); sandor.daranyi@hb.se (S. Darányi)
0009-0007-6621-9696 (T. van Erven); 0009-0001-7215-8582 (S. Kontogiannis); 0000-0002-4435-2521 (E. Kontopoulos); 0000-0002-4873-5255 (K. Avgerinakis); 0000-0001-8879-8798 (A. Scharnhorst); 0000-0002-1542-934X (S. Darányi)

embrace the shift of concepts has never been far away [1]. Large scale data visualisations (often as networks) have been used to display the evolution of ideas, cultural artefacts, and in essence any traces of human activity [2], [3]. Digital humanities and cultural heritage collections have attracted specific interest because they are so deeply connected to what defines humanity and society [4]. But there remains one question: can we also understand the underlying principles, processes, laws which drive this evolution? We don't claim to have an ultimate answer or solution but claim to have found a way to develop an alternative heuristic approach to think about knowledge dynamics.

This paper has been submitted to the *Blue Sky* track because it is cross-disciplinary. Conceptually, our approach is rooted in complexity science applied to knowledge dynamics. We build on decades of digitisation of artefacts of human creativity and of fine-tuning the way those artefacts are represented in metadata schemas. In the core of this paper we describe a workflow called *Knowledge Tectonics* (KT), in which data collection and curation and state of the art in knowledge engineering design come- together in a research-engineered interface, enabling us to visually analyse trajectories of objects, and for the first time to also visualise *forces* behind those trajectories.

We use as a proof of concept a standard collection of famous paintings, in which content clustering appears as the temporal "lithosphere" for scalable multi-century analysis, velocity vector fields and streamlines show stylistic migration, and the edges of a Vector Augmented Knowledge Graph (VAKG) introduce a kind of viscosity that resists possible processes of feature drift or epistemic change. The benefit of this model is its alignment with important past work on information visualization, a predecessor of knowledge emergence as semantic plate tectonics [5]-[9]. At the same time, our proposal departs from a previous attempt to model content dynamics as a series of Newtonian gravitational potential maps [10]. Namely, gravitational paradigms are fundamentally limited: they only capture static draw, causing entities to permanently collapse into local minima. In contrast, continuum geodynamics introduces **viscosity** and **rotational curls**. Human inquiry does not simply drop into static buckets; it exhibits structural resistance to modification (viscosity), linear innovation vectors (divergence), and cyclical stylistic revivals (solenoidal fields). Fluid dynamics provides the explicit partial differential calculus needed to separate these distinct behavioral layers.

## 2. Data

We use the WikiArt Emotions dataset consisting of 4,105 paintings spanning 600 years [11]. The fact that this dataset combines artworks, temporal coverage, stylistic labels, image material, and affective descriptors, makes it suitable for testing whether the model of KT can reveal how cultural and semantic formations move, cluster, and stabilize across historical time. The dataset was initially available as JSON metadata linked to image files, including ID, title, artist, production year, style/category, image URL, PageRank, and affective scores [10].

Related WikiArt-derived KG work already exists, most notably ArtGraph, which integrates WikiArt and DBpedia plus provides an artistic KG in RDF form [12]. However, our goal is different. We do not present the transformation of WikiArt or WikiArt Emotions into RDF as the main contribution of this paper. Instead, we construct a task-specific RDF representation as a semantic substrate for KT. This representation includes artwork entities, artists, movements, periods, style categories, affective descriptors, image references, and generated textual annotations, because these relations are needed to derive KG embeddings and graph-based signals for the tectonic model. In future work, this RDF layer could be aligned with ArtGraph or other Linked Open Data resources to enrich its contextual grounding.

## 3. Methodology – The Knowledge Tectonics Workflow

We assume that the production of artistic artefacts, as well as the production of other artefacts of human ingenuity, are governed by similar dynamics, hence the use of the above label as an umbrella term for our approach. The workflow contains four steps through which existing metadata and information from the images themselves are transformed into a graph-augmented semantic landscape, where image embeddings define positions and signals derived from KGs define pressure, viscosity, and drift in a way detailed later. Figure 1 gives a schematic overview about those stages.

During Stage 1 we executed a couple of data curating steps to make the dataset ready for the KT workflow. The image file for each artwork was used to produce rich caption embeddings using a LLaVA model [13] from the normalized image files, which were first reshaped to optimal dimensions using Python's Pillow library [14]. Based on the AI-generated rich captions, SentiArt [15] was then deployed for computing the Average Affective Potential (AAP). Finally, PageRank scores were computed based on the captions using NetworkX [16]. Graph nodes represented Artwork, Artist, Style Category and Period. Stage 2 was devoted to KG embedding (KGE) construction, isolating the components, and downstream use of KGE vectors. For this stage, we used PyKEEN, a Python library for training and evaluating knowledge graph embedding models, to generate the KGE vectors used in the downstream tectonic workflow [17].

Stage 3 brings the dimension of time into the workflow. Uniform Manifold Approximation and Projection (UMAP) is a dimension reduction technique that can be used for visualisation of higher-dimensional objects (such as networks). It is known to enable the projection of complex, high-dimensional datasets into 2D or 3D spaces while preserving both local and global data structures [18]. RDF graphs are formally atemporal and are often treated as static snapshots of information, even though RDF datasets and named graphs can manage multiple contexts or versions [19]. While this is valuable for integration, interoperability, querying, and retrieval, it is less suited to modelling how semantic formations move, stabilize, fragment, or reappear over time.

Existing temporal KG approaches and graph evolution models [20]-[22] mainly focus on changing facts, temporal representation learning, KG completion, and link prediction, rather than semantic pressure, resistance, and directional drift as field-like phenomena. While vector-space and KG embedding methods capture similarity patterns in low-dimensional spaces, they reduce the direct ability to inspect typed relations, provenance and conceptual structure found in RDF/OWL representations. Our workflow combines temporal KG slicing with a focus on the connections between time slices.

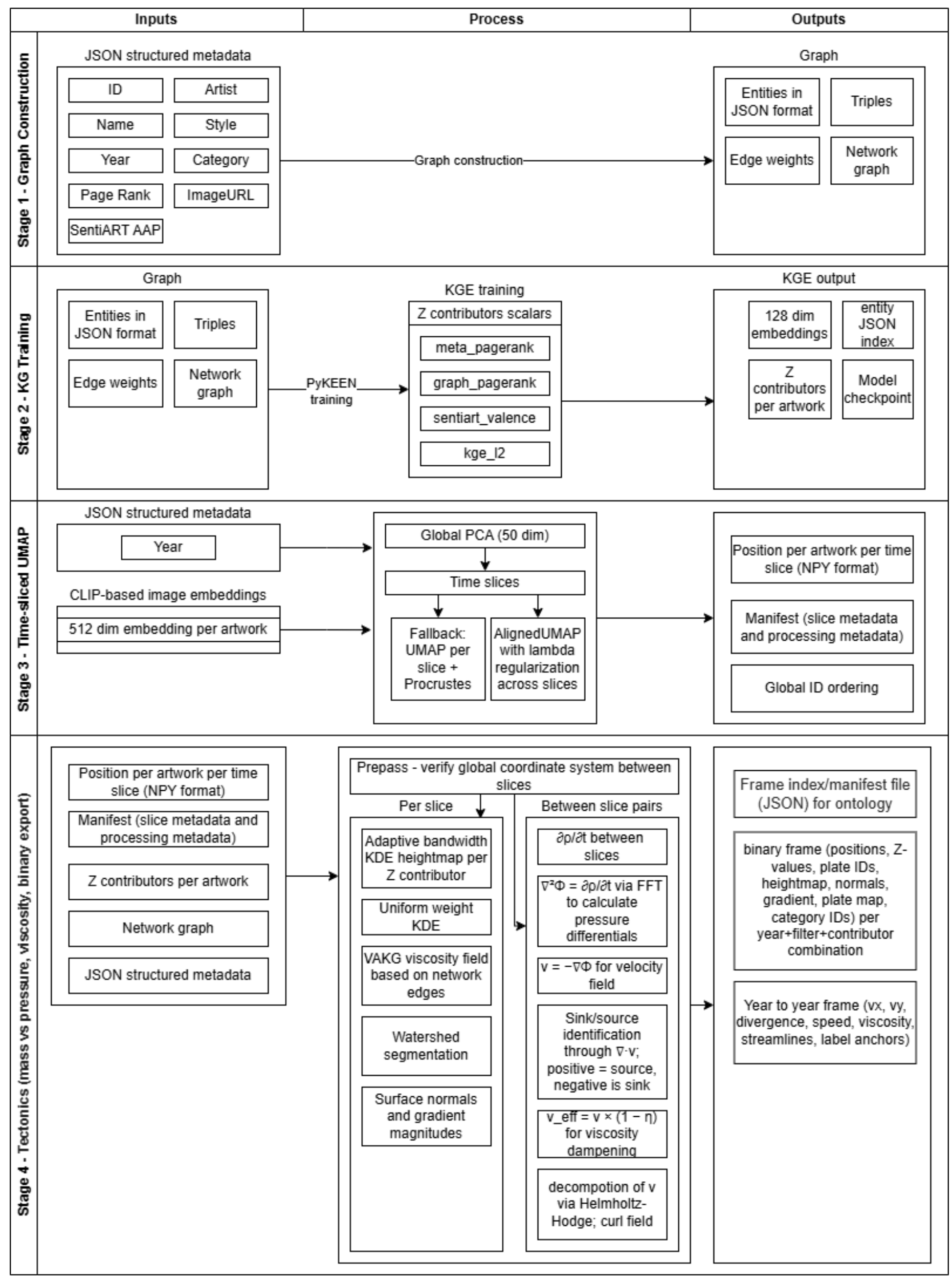

**Figure 1:** KT workflow. Arrows indicate processing transformations between workflow stages, not RDF predicates, KG edges, or causal relations. Metadata and images are first converted into typed graph and embedding representations; these are then temporally sliced and used to compute density, pressure, viscosity, velocity, divergence, and HHD-based drift indicators.

In the RDF substrate, time is represented first as artwork-level production metadata, while the temporal slices used by KT are model-level constructs rather than intrinsic historical periods. A slice therefore has to be treated as a provenance-bearing analytical view over the assertion graph: it records the interval boundaries, granularity choice, embedding model, KDE bandwidth, and graph-weighting configuration used for a given KT run. This distinction prevents the KG from conflating cultural-historical assertions, such as the production year of an artwork, with model-derived temporal groupings used to compute pressure, viscosity, and drift.

Step 4 is the core of the KT workflow, showing how new nodes and changing relations reposition entities over time, revealing continuities and hidden patterns of relatedness. Similar dynamics appear in science as schools of thought or phylogeny [23], and in art history as schools, genres, traditions and paradigms persist, break, or emerge [4]. To visualise these forces, several data processing operations are applied, whose mathematical definitions and workflow roles are outlined below and summarised in Figure 1.

### 3.1. Kernel Density Estimation (KDE)

To turn discrete artwork coordinates into a continuous "epistemic lithosphere," we used an Adaptive KDE to compensate for a tail-weighted dataset. Unlike traditional histograms, this approach treats each artwork as a localized "semantic mass" that contributes to a global density field, defined by:

$$\rho(x, y) = \sum_{i=1}^{N} w_i \cdot \frac{1}{2\pi h_i^2} \exp\left(-\frac{(x - x_i)^2 + (y - y_i)^2}{2h_i^2}\right) \tag{1}$$

Here, the bandwidth $h_i$ is determined by the $k$-th nearest neighbor, allowing the model to adapt to both dense stylistic clusters and isolated experimental "islands".

### 3.2. VAKG Viscosity Field

The Vector-Augmented Knowledge Graph provides the structural resistance to drift. We first calculate the normalized weighted degree for each artwork:

$$w_i = \frac{\sum_j W_{ij}}{\max_k \sum_j W_{kj}} \tag{2}$$

The viscosity field $\eta(x, y)$ is then derived as a smoothed Gaussian KDE (sigma = 2.0) over these weights. The final effective velocity$v_{eff}$ accounts for this damping:

$$\mathrm{v}_{eff} = v \cdot (1 - \eta) \tag{3}$$

The viscosity field is derived from the KG edge weights. Because this structural resistance is derived entirely from ontological connections, the resulting damping effect is purely semantic—a characteristic that distinguishes our geodynamic model from a conventional physics simulation.

### 3.3. The Poisson Pressure Solve

We model stylistic change as a pressure potential $\Phi$ generated by the density flux between time slices (sparse periods = 25 years, dense periods = 10 years, both normalized by time)

The Poisson Equation (Eq. 4) is the mathematical link between "information change" ($\partial\rho/\ \partial t$) and "semantic pressure" ($\Phi$):

$$\nabla^2\Phi = \frac{\partial\rho}{\partial t} \approx \rho_{\text{current}} - \rho_{\text{previous}} \quad (4)$$

To solve this efficiently, we move to frequency space using a Fourier Transform (where $k$ is the wavenumber vector):

$$\widehat{\Phi}(k) = -\frac{\widehat{\partial\rho/\partial t(k)}}{|k|^2} \quad (5)$$

The choice of temporal slicing acts as a grid-resolution filter on the underlying partial differential equations. In our architecture, setting sparse intervals to 25 years and dense eras to 10 years prevents numerical instabilities. Slicing too narrowly ($\Delta(t) < 5$ years) creates localized noise due to sparse data distribution, causing erratic pressure spikes. Conversely, over-aggregating ($\Delta(t) > 50$ years) leads to temporal aliasing, smoothing out vital, high-frequency ideological shifts. Aligning this with temporal graph designs (e.g., [20]) highlights how geodynamics can naturally balance data sparsity against temporal continuity.

### Velocity and Divergence

The velocity field $\boldsymbol{v}$ (the "drift") is the negative gradient of this potential:

$$v = -\nabla\Phi = \left(-\frac{\partial\Phi}{\partial x}, -\frac{\partial\Phi}{\partial y}\right) \quad (6)$$

The divergence identifies semantic sources (protrusions) and sinks (coalescence):

$$\nabla \cdot v = \frac{\partial v_x}{\partial x} + \frac{\partial v_y}{\partial y} \quad (7)$$

Divergence ($\nabla \cdot v$) is directly linked to visualization: a "source" (positive divergence) represents a location where new ideas are actively dispersing, while a "sink" (negative divergence) represents where styles are coalescing or becoming conventional.

## 3.4. Helmholtz-Hodge Decomposition (HHD)

To separate direct innovation from stylistic cycles, we decompose the field into **irrotational** (potential-driven) and **solenoidal** (rotational) components:

$$v = \nabla\phi + \nabla \times A \quad (8)$$

In the frequency domain, the **irrotational component** is isolated as:

$$\hat{\mathrm{v}}_{irrot}(k) = \frac{\hat{v}(k) \cdot k}{|k|^2} k \tag{9}$$

The **solenoidal fraction** ($f_{\mathrm{sol}}$) serves as our core metric for quantifying the "cyclicity" versus "linearity" of knowledge drift:

$$f_{sol} = \frac{\langle |\mathrm{v}_{sol}|^2 \rangle}{\langle |\mathrm{v}_{irrot}|^2 \rangle + \langle |\mathrm{v}_{sol}|^2 \rangle} \tag{10}$$

These equations can be interpreted physically as follows: the Poisson equation explains movement through pressure, viscosity captures resistance from graph structure, and HHD distinguishes between emerging trends and revivals. The resulting images from our Python workflow (Figure 1) combine 2D image embeddings (512-d CLIP [24]) for object positioning with KG embeddings derived from RDF graphs built from the original JSON metadata, encoding height, viscosity, and L2 norms. RDF serves as the semantic layer providing explicit structure, typed relations, and provenance, extending beyond "flat" distributional vector spaces toward a predictive model of cultural geodynamics.

## 4. Typical Results, Evaluation and Discussion

An interactive tool (available at https://tectonics.wizardtower.dev) was built based on the KT workflow. As per the mathematical framework introduced above, information visualization is used to demonstrate dynamics and evaluated by an extension of knowledge-graph semantics. Once selected KT observations are materialized in a derived observation graph, the workflow supports SPARQL queries over velocity, viscosity, divergence, and solenoidal-fraction combinations, e.g., querying whether an emergent school of painting met graph-derived structural resistance in a particular period. Figure 2 shows typical results from the interface.

While the KT framework is novel, it maps directly onto several established fields. To measure the "system dynamics" of content drift as a new class of metadata, one can look to three specific domains: Quantitative System Dynamics, Scientometrics/Bibliometrics, and Data Stream Mining. Recommended change parameters include velocity and momentum for the rate of shift by expansion rate/dilation, semantic velocity ($v$), and inertia; semantic drift and concept drift metrics by Kullback-Leibler divergence and semantic longevity, the "half-life" of a cluster before its connectivity (VAKG edges) or position significantly degrades or shifts; and Helmholtz-Hodge Metrics by solenoidal fraction ($f_{sol}$) for a "cyclicity index" measuring recycled approaches as fashion vs. innovation, and vorticity/curl ($\omega$), measuring the intellectual "turbulence" of a specific period.

A typical finding would be the following: "The 1945–1955 transition exhibited high expansion rate (+15) in the Abstract cluster, characterized by a low solenoidal fraction, indicating a period of linear innovation rather than stylistic revival." This statement should be read as a model-derived diagnostic, not as an art-historical conclusion by itself. In this case, "high expansion rate" means that the local density of Abstract-labelled works increases and spreads between temporal slices, while "low solenoidal fraction" means that the inferred field is dominated by directional expansion rather than rotational or revival-like movement. The claim therefore identifies a historically interesting period/cluster pair for expert interpretation, rather than proving that Abstract art evolved linearly in the historical sense.

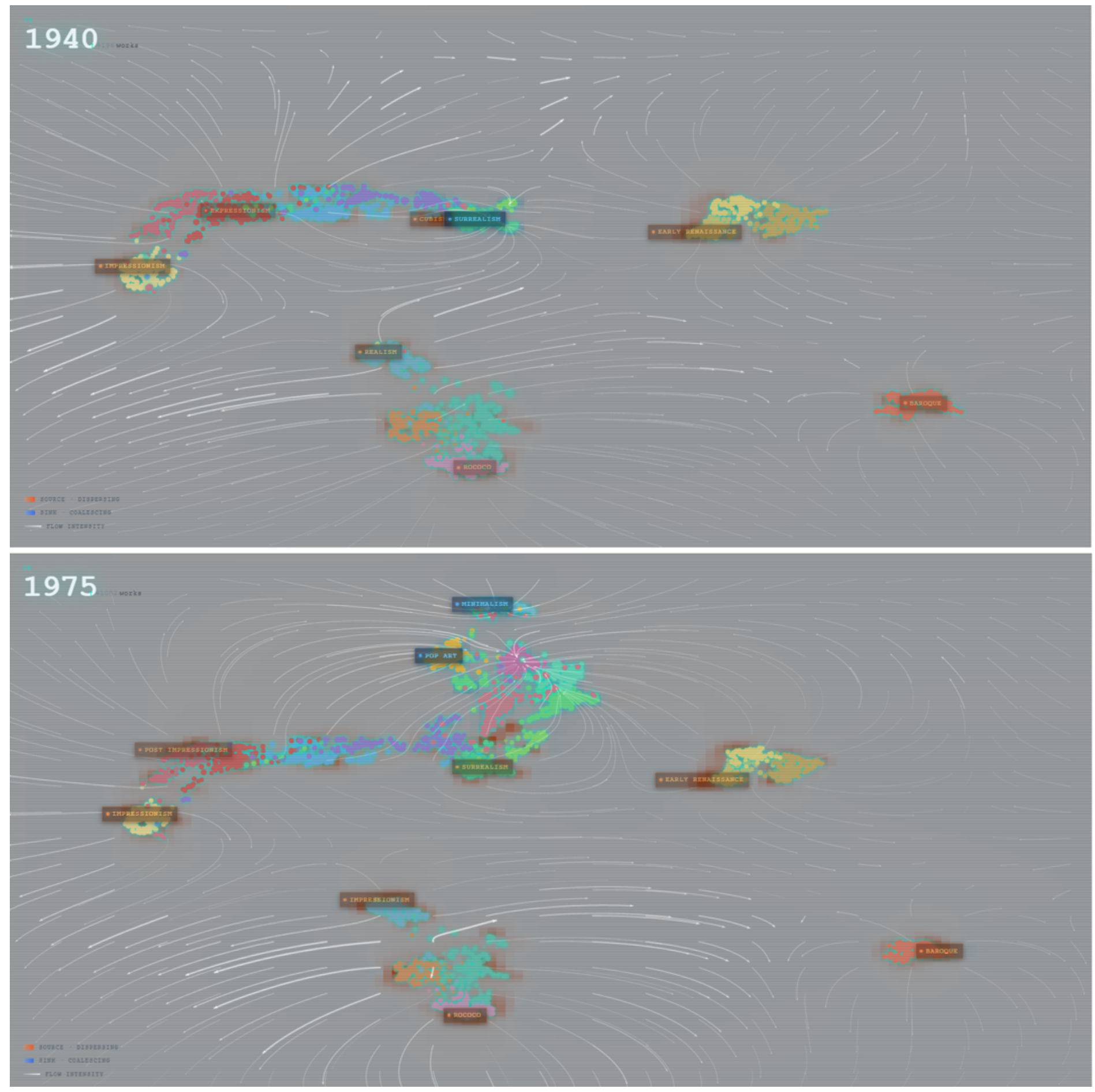

**Figure 2:** Vector field modelling of semantic pressure across historical slices (1940 vs. 1975). Lines indicate the movement of artwork groups over time, while dense areas mark emerging or stable styles and emotions. Brown and blue zones represent high and low semantic pressure, respectively. The white background streamlines do not represent the migration lines of individual artworks; rather, they visualize the underlying vector field forces guiding the overall cultural flow, akin to wind currents on a weather map.

KT bridges traditional and new metadata approaches by combining RDF-based semantics with vector and field-based modelling. In this proof of concept, RDF provides the typed semantic substrate for deriving graph features and KG embeddings, while vector and field methods model drift across a continuous landscape. Artwork entities are linked to artists, movements, periods, concepts, affective labels, and image-derived annotations, whose graph-based signals shape the viscosity field. While KT outputs such as pressure zones, source/sink regions, and drift events do not yet materialize back into RDF, this is a natural next step for enabling queryable and reusable drift patterns.

While materializing the continuous vector fields directly into RDF triples would cause severe graph bloat, materializing the discrete *outputs* of these fields is highly desirable. Using RDF-star [25], macro-level tectonic transformations can be explicitly asserted. For example, an artistic movement can be decorated with a *solenoidalFraction* or a specific historical window can be asserted as an *orogenyEvent* (conceptual collision). This allows standard SPARQL engines to index and query systemic drift without storing raw coordinate mechanics.

Content behaviour under intellectual pressure can be compared to geological transformations such as subduction, collision, and divergence. By projecting data onto a dynamic manifold and calculating flow differentials across time slices, the framework visualizes these cultural transformations as measurable physical phenomena.

## 5. Future Research

We offered evidence for content migration in art history modelled as streamlines with curls in expanding vector fields, always aligned with their evolving KG states. Whereas data science typically explains phenomena through statistical correlation and significance, our approach models the data *as if* generated by an underlying dynamical system, borrowing the formalism of classical mechanics. This generative, physics-inspired stance is a genuinely novel aspect of our work.

A next, more detailed paper will address evaluation, plus introduce probabilistic Langevin flows on scalable citation datasets to demonstrate citation counts as sediment flux eroding or building the semantic landscape.

## 6. Towards Cultural Geodynamics: Vision and Horizons

As a vision statement, we have formalized KT, a paradigm shift that addresses a long-standing challenge in the Semantic Web: the lack of a continuous, physically grounded model for tracking conceptual transformation. By embedding the symbolic structural constraints of RDF graphs into the continuous field equations of fluid mechanics, we couple static vector representations with graph-based semantic structure to capture the deterministic rules of *epistemic changes* (concept and epistemic drift, phylogenetics). While validated here on centuries of artistic evolution, the long-term implications of KT extend far beyond digital humanities. By measuring knowledge through a unified lexicon of semantic velocity, pressure gradients, and solenoidal fractions, this framework provides a cross-disciplinary diagnostic engine applicable to tracking scientific innovation, monitoring geopolitical narrative shifts, and managing enterprise knowledge quality. Ultimately, as AI-driven information environments continue to accelerate data churn, KT offers a vital methodology for quantifying, predicting, and safeguarding the structural integrity of human inquiry across generations.

## Acknowledgements

Respective concepts and their partial applications were worked out in the SILVANUS (Grant agreement ID: 101037247) and MuseIT (Grant agreement ID: 101061441) projects. Stelios Kontogiannis and Kostas Avgerinakis contributed by means of the EXCALIBUR project (Grant Agreement ID: 101233712).

**Declaration on Generative AI.** *During the preparation of this work, the authors used Gemini 3.0 in order to: Grammar and spelling check, Paraphrase and reword. After using this tool/service, the authors reviewed and edited the content as needed and take full responsibility for the publication's content.* **Cf. https://ceur-ws.org/GenAI/Taxonomy.html**